\documentstyle[aps]{revtex}

\begin{document}
\title{Time Dependent Analysis of Decays $\Lambda _b\rightarrow \Lambda +D^0$ and $%
\Lambda _b\rightarrow \Lambda +\bar{D}^0$ }
\author{Fayyazuddin}
\address{National Centre for Physics and Physics Department, Quaid-i-Azam University,%
\\
Islamabad, Pakistan.\\
submitted to Phys. Lett. B}
\maketitle

\begin{abstract}
The time-dependent analysis of the decays $\Lambda _b\rightarrow \Lambda
+D^0(t)$ and $\Lambda _b\rightarrow \Lambda +\bar{D}^0(t)$ is discussed. The
effect of particle mixing due to time evolution of $D^0$ and $\bar{D}^0$ on
the observables for these decays viz the branching ratio of decay widths and
the asymmetry parameters $\alpha ,\beta $ and $\gamma $ are analysed. It is
shown that it is possible to extract information about $(\Delta m/\Gamma
)\sin \hat{\gamma}$ from the experimental data for these observables. Here $%
\Delta m=m_{D_1^0}-m_{D_2^0},\Gamma $ is the decay widths for $D^{\prime }s$
and $\hat{\gamma}$ is the weak phase.
\end{abstract}

In this paper, we discuss the time dependent analysis of the decays $\Lambda
_b\rightarrow \Lambda +D^0(t)$ and $\Lambda _b\rightarrow \Lambda +\bar{D}%
^0(t)$. Due to $D^0$ and $\bar{D}^0$ mixing, a pure $D^0(\bar{D}^0)$ beam
acquires a component of $\bar{D}^0(D^0)$ as it evolves. We analyse the
effect of particle mixing on the diservables for these decays viz the ratio
of decay widths and the asymmetry parameters, $\alpha ,\beta ,$ and $\gamma
. $

These decays have been previously studied in references $[1]$ and $[2]$;
especially the decays $\Lambda _b\rightarrow \Lambda +D_{1,2}$ where $D_{1,2}
$ are CP-eigenstates. The decays are described by four amplitudes $A_D(t),A_{%
\bar{D}}(t),B_D(t)$ and $B_{\bar{D}}(t)$. Denoting these amplitudes as $%
R_D(t)$ and $R_{\bar{D}}(t),$ where $R=A$ or $B$, we get the time dependent
amplitudes.
\begin{eqnarray}
\left| R_D(t)\right| ^2 &=&\frac 12e^{-\Gamma t}\left[ (1+\cos \Delta
mt)R_D^2-2\sin \Delta mt\sin \hat{\gamma}R_DR_{\bar{D}}+(1-\cos \Delta
mt)R_D^2\right]  \\
\mathop{\rm Re}
A_D^{*}(t)B_D(t) &=&\frac 12e^{-\Gamma t}\left[ (1+\cos \Delta
mt)A_DB_D+\sin \hat{\gamma}\sin \Delta mt\text{ }(A_DB_{\bar{D}}+A_{\bar{D}%
}B_D)+(1-\cos \Delta mt)A_{\bar{D}}B_{\bar{D}}\right]  \\
\mathop{\rm Im}
A_D^{*}(t)B_D(t) &=&\frac 12e^{-\Gamma t}[(1+\cos \hat{\gamma}\sin \Delta
mt)(A_{\bar{D}}B_D-A_DB_{\bar{D}})
\end{eqnarray}
where we have explicitly exhibited the weak phase $\hat{\gamma}$. After
taking out the weak phase, these amplitudes are real, if we neglect the
final state interactions. For $\bar{D},$ change $D\rightarrow \bar{D}$ and $%
\sin \Delta mt\rightarrow -\sin \Delta mt$ in Eqs. $(1),(2)$ and $(3).$ We
now take the time average of these amplitudes:

\begin{equation}
\bar{R}_D^2=\frac{\int_0^\infty R_D^2(t)dt}{\int_0^\infty e^{-\Gamma t}dt}
\end{equation}

After taking the time average and neglecting terms of the order $(\Delta
m/\Gamma )^2,$ we obtain (similar results follow if instead of taking time
average, we take $t\sim \frac 1\Gamma )$

\begin{eqnarray}
\bar{R}_D^2 &=&\approx \left[ R_D^2+(\Delta m/\Gamma )\sin \hat{\gamma}R_DR_{%
\bar{D}}\right] \\
\bar{\alpha}_D &=&2\left| \vec{k}\right| \left[ A_DB_D+\frac 12(\Delta
m/\Gamma )\sin \hat{\gamma}A_DB_{\bar{D}}+A_{\bar{D}}B_D)\right] /\bar{F}_D^2
\\
\bar{\beta}_D &=&2\left| \vec{k}\right| \left( \Delta m/\Gamma \right) \cos 
\hat{\gamma}\frac 12\left( A_{\bar{D}}B_{\bar{D}}-A_DB_{\bar{D}}\right) /%
\bar{F}_D^2 \\
\bar{\gamma}_D &=&\frac{\left[ (E_{\wedge }+m_{\wedge })A_D^2-(E_{\wedge
}-m_{\wedge })B_D^2\right] +(\Delta m/\Gamma )\sin \hat{\gamma}\left[
(E_{\wedge }+m_{\wedge })A_DA_{\bar{D}}-(E_{\wedge }-m_{\wedge })B_DB_{\bar{D%
}}\right] }{\bar{F}_D^2}
\end{eqnarray}
where

\begin{eqnarray}
\bar{F}_D^2 &=&(E_{\wedge }+m_{\wedge })\bar{A}_D^2+(E_{\wedge }-m_{\wedge })%
\bar{B}_D^2  \nonumber \\
&=&\left[ (E_{\wedge }+m_{\wedge })A_D^2+(E_{\wedge }-m_{\wedge
})B_D^2\right]  \nonumber \\
&&+(\Delta m/\Gamma )\sin \hat{\gamma}\left[ (E_{\wedge }+m_{\wedge })A_DA_{%
\bar{D}}+(E_{\wedge }-m_{\wedge })B_DB_{\bar{D}}\right]
\end{eqnarray}

For $\bar{D},$ change $(\Delta m/\Gamma )$ to - $(\Delta m/\Gamma )$ and $%
A_D,B_D\leftrightarrow A_{\bar{D}},B_{\bar{D}}$ in $Eqs.(5),(6),(7),(8)$ and 
$(9).$

It is convenient to put
\begin{equation}
A_D=\frac{a_D}{\sqrt{E_{\wedge }+m_{\wedge }}},B_D=\frac{b_D}{\sqrt{%
E_{\wedge }-m_{\wedge }}}
\end{equation}

In terms of these amplitudes, we have from Eqs. $(6-9)$,
\begin{eqnarray}
\bar{F}_D^2 &=&(a_D^2+b_D^2)\left[ 1+(\Delta m/\Gamma )\sin \hat{\gamma}%
\frac{a_Da_{\bar{D}}+b_Db_{\bar{D}}}{a_D^2a_{\bar{D}}^2}\right]  \\
\bar{\alpha}_D &=&\frac{2a_Db_D}{a_D^2+b_{\bar{D}}^2}\left[ 1+(\Delta
m/\Gamma )\sin \hat{\gamma}\left( \frac 12(\frac{b_{\bar{D}}}{b_D}+\frac{a_{%
\bar{D}}}{a_D})-\frac{a_Da_{\bar{D}}+b_Db_{\bar{D}}}{a_D^2+a_{\bar{D}}^2}%
\right) \right]  \\
\bar{\beta}_D &=&(\Delta m/\Gamma )\cos \hat{\gamma}\frac{a_{\bar{D}%
}b_D-a_Db_{\bar{D}}}{a_D^2+b_D^2} \\
\bar{\gamma}_D &=&\frac 1{a_D^2+b_D^2}\left[ \left( a_D^2-b_D^2\right)
+\left( \Delta m/\Gamma \right) \sin \hat{\gamma}\left( \left( a_Da_{\bar{D}%
}+b_Db_{\bar{D}}\right) -\left( a_D^2-b_D^2\right) \frac{\left(
a_D^2-b_D^2\right) \left( a_Da_{\bar{D}}+b_Db_{\bar{D}}\right) }{a_D^2+b_D^2}%
\right) \right] 
\end{eqnarray}

For $\bar{D}$, change $(\Delta m/\Gamma )\rightarrow -(\Delta m/\Gamma ),$ $%
a_D,b_D\leftrightarrow a_{\bar{D}},b_{\bar{D}}$ in Eqs. $(11),(12),(13)$ and 
$\left( 14\right) .$

To proceed further, we note that in the factorization ansatz $[2]$

\begin{eqnarray}
a_{\bar{D}} &=&\frac{|V_{ub}||V_{cs}|}{|V_{cb}||V_{us}|}a_D\simeq \sqrt{\rho
^2+\eta ^2}a_D  \nonumber \\
b_{\bar{D}} &=&\sqrt{\rho ^2+\eta ^2}\frac{b_D}{1+x}. \\
a_D &=&-\frac{G_F}{\sqrt{2}}|V_{cb}V_{us}|a_2F_D(m_{\wedge _b}-m_{\wedge
})g_V\sqrt{E_{\wedge }+m_{\wedge }} \\
b_D &=&\frac{G_F}{\sqrt{2}}|V_{cb}V_{us}|a_2F_D(m_{\wedge b}+m_{\wedge })g_A%
\sqrt{E_{\wedge }-m_{\wedge }}(1+x)
\end{eqnarray}

Here $x=\frac{b_p}{b_f}$ and $b_p$ is the baryon poles contribution which
contributes only to $b_D.b_f=a_2F_D(m_{\wedge b}+m_{\wedge })g_A.$ Note that
in Eq. $(15),$ we have used Wolfenstein parametrization $[3]$ of CKM matrix $%
[4]$. We will take $g_V=g_A$. Thus we can write

\begin{equation}
b_{\bar{D}}/a_{\bar{D}}=-d,\text{ }b_D/a_D=-d(1+x)
\end{equation}

where

\begin{equation}
d=\frac{m_{\wedge _b}+m_{\wedge }}{m_{\wedge _b}-m_{\wedge }}\sqrt{\frac{%
m_{\wedge b}-m_{\wedge }}{m_{\wedge b}+m_{\wedge }}}\simeq 0.946
\end{equation}

on using $m_{\wedge _b}=5.624GeV$ and $m_{\wedge }=1.116GeV.$

Using Eqs. $(18)$ and $(15)$, we obtain from Eqs. $(11),(12),(13)$ and $%
(14). $

\begin{eqnarray}
\delta  &\equiv &\frac{\Gamma (\wedge _b\rightarrow \wedge +\bar{D}^0)}{%
(\rho ^2+\eta ^2)\Gamma (\wedge _b\rightarrow \wedge +D^0)}=\frac 1{(\rho
^2+\eta ^2)}\frac{\bar{F}_{\bar{D}}^2}{\bar{F}_D^2}  \nonumber \\
&=&\frac{1+d^2}{1+d^2(1+x)^2}\left[ 1-\left( \Delta m/\Gamma \right) \sin 
\hat{\gamma}\frac 1{\sqrt{\rho ^2+\eta ^2}}\left( 1+\frac{d^2}{1+d^2}%
x\right) \right]  \\
\bar{\alpha}_D &=&\frac{-2d(1+x)}{1+d^2(1+x)^2}\left[ 1-\sqrt{\rho ^2+\eta ^2%
}\left( \Delta m/\Gamma \right) \sin \hat{\gamma}\left( \frac x{2(1+x)}%
\right) \frac{1-d^2(1+x)^2}{1+d^2(1+x)^2}\right]  \\
\bar{\beta}_D &=&-\sqrt{\rho ^2+\eta ^2}\left( \Delta m/\Gamma \right) \cos 
\hat{\gamma}\frac{dx}{1+d^2(1+x)^2} \\
\bar{\gamma}_D &=&\frac 1{1+d^2(1+x)^2}\left[ 1-d^2(1+x)^2+\sqrt{\rho
^2+\eta ^2}\left( \Delta m/\Gamma \right) \sin \hat{\gamma}\frac{2d^2x(1+x)}{%
1+d^2(1+x)^2}\right]  \\
\bar{\alpha}_{\bar{D}} &=&\frac{-2d}{1+d^2}\left[ 1-\left( \Delta m/\Gamma
\right) \sin \hat{\gamma}\frac 1{\rho ^2+\eta ^2}\frac{x(1-d^2)}{2(1+d^2)}%
\right]  \\
\bar{\beta}_{\bar{D}} &=&-\frac{\left( \Delta m/\Gamma \right) \cos \hat{%
\gamma}}{\sqrt{\rho ^2+\eta ^2}}\text{ }\frac{dx}{1+d^2} \\
\bar{\gamma}_{\bar{D}} &=&\frac{1-d^2}{1+d^2}+\left[ 1+\frac 1{\sqrt{\rho
^2+\eta ^2}}\left( \Delta m/\Gamma \right) \sin \hat{\gamma}\frac{2d^2x}{%
(1-d^4)}\right] 
\end{eqnarray}

We first note that the inteference effect is more pronounced in the
observables for $\bar{D}$, since the admixture of D tends to enhance it by a
factor$\frac 1{\sqrt{\rho ^2+\eta ^2}}$. But since $d^2=0.895,$ the
interference effect in $\bar{\alpha}_{\bar{D}}$ is neglegible. Also we note
that this effect venishes in $\alpha ,$ $\beta $ and $\gamma $ for $x=0$.
But there is no reason to believe that $x=0$ as shown in reference $[2].$
Just to give an estimate of the effect of $D^0\rightarrow \bar{D}^0$ mixing,
using $x=-0.64$ $[2],$ $(\rho ,\eta )=(0.05,$ $0.36)$ $[5],$ we get

\begin{eqnarray}
1.38 &\leq &\delta \leq 2.03  \nonumber \\
0.006 &\leq &\bar{\gamma}_{\bar{D}}\leq 0.104
\end{eqnarray}

Without interference effect $\delta =1.70$ and $\bar{\gamma}_{\bar{D}%
}=0.055. $ In deriving the inequality $(27)$, we have used the experimental $%
[6]$ upper limit $|\Delta m/\Gamma |<0.10.$

First we note that asymmetry parameter $\beta $ which characterizes
CP-violation is a consequence of particle mixing. The experimental upper
limit on $|\Delta m/\Gamma |$ gives

\[
-0.1\leq (\Delta m/\Gamma )\sin \bar{\gamma}\leq 0.1 
\]

Thus if we plot $\delta $ and $\bar{\gamma}_{\bar{D}}$ as function of $%
(\Delta m/\Gamma )\sin \bar{\gamma}$ in the range $-0.1$ to $0.1$, treating $%
x$ as a free parameter lying in the range $-1<x<1$, we may be able to
extract some information for $(\Delta m/\Gamma )\sin \bar{\gamma}$ from the
experimental values of $\delta $ and $\bar{\gamma}_{\bar{D}}$.
Experimentally, measurement of the ratio $\delta $ should not be very
difficult. In Figs. $1$ and $2$ we have plotted $\delta $ and $\bar{\gamma}_{%
\bar{D}}$ vs $(\Delta m/\Gamma )\sin \bar{\gamma}$ for the four values of $x$
viz $x=-0.8,-0.6,-0.4,0.4,0.6$ and $0.8$ taking $\sqrt{\delta ^2+\eta ^2}%
=0.36.$ If $\sin \hat{\gamma}$ is too small, then $(\Delta m/\Gamma )\cos 
\hat{\gamma}$ may be extracted form similar plot for $\bar{\beta}_{\bar{D}}$
as shown in $Fig.3.$

To conclude the mixing of $D^0-\bar{D}^0$ has some observable effects in the
decays $\wedge _b\rightarrow \wedge +D^0$ and $\wedge _b\rightarrow \wedge +%
\bar{D}^0$. The branching ratio $\delta $ for these decays can give some
information to extract the value of $(\Delta m/\Gamma )\sin \hat{\gamma},$
provided $\sin \hat{\gamma}$ is not too small.

\begin{center}
{\bf Acknowledgement}
\end{center}

I am grateful to Prof. Riazuddin for helpfull discussions.

\begin{center}
{\bf Figure Captions}
\end{center}

\begin{enumerate}
\item  Plot of $\delta $ vs $(\Delta m/\Gamma )\sin \hat{\gamma}$ (cf. Eq.
20) for $x=-0.8,-0.6,-0.4,0.4,0.6$ and $0.8.$

\item  Plot of $\bar{\gamma}_{\bar{D}}$ vs $(\Delta m/\Gamma )\sin \hat{%
\gamma}$ (cf. Eq. 26) for $x=-0.8,-0.6,-0.4,0.4,0.6$ and $0.8.$

\item  Plot of $\bar{\beta}_{\bar{D}}$ vs $(\Delta m/\Gamma )\cos \hat{\gamma%
}$ (cf. Eq. 25) for $x=-0.8,-0.6,-0.4,0.4,0.6$ and $0.8.$
\end{enumerate}

\end{document}